\documentclass
{aa}

\usepackage{aabib99}
\usepackage{graphicx}
\usepackage[american]{babel}

\begin{document}

\title{Keplerian frequencies and innermost stable circular orbits of rapidly
rotating strange stars
}

\titlerunning{Keplerian frequencies of strange stars}

\author{ 
   N. Stergioulas$^{1,2}$,
   W. Klu\'zniak$^{3}$,
   T. Bulik$^3$,
  }

\authorrunning{ Stergioulas,  Klu\'zniak, Bulik }

\institute{$^1$ Max-Planck-Institut f{\"u}r Gravitationsphysik,
 Am M{\"u}hlenberg 5, D-14476 Golm, Germany\\
$^2$ Department of Physics, Aristotle University of Thessaloniki,
Thessaloniki 54006, Greece \\
$^3$Nicolaus Copernicus Astronomical Center, 
Bartycka 18, 00-716 Warszawa, Poland}

\date{Received , Accepted }

\thesaurus{02.04.01, 08.14.1}

\maketitle

\begin{abstract}

It has been suggested that the frequency in the co-rotating innermost
stable circular orbit (ISCO) about a compact stellar remnant
can be determined through X-ray observations of low-mass X-ray binaries,
and that its value can be used to constrain the equation of state
of ultradense matter.
 Upon constructing numerical models of rapidly rotating strange
(quark) stars in general relativity, we find that for stars rotating
at the equatorial mass-shedding limit, 
the ISCO is indeed above the stellar surface,
for a wide range of central energy densities at a height
equal to 11\% of the circumferential stellar radius, which scales inversely
with the square root of the energy density, $\rho_0$,
of self-bound quark matter at zero presure. For these models, the ISCO
frequency is $81.5\pm1.5$\% of the stellar rotational frequency,
whose maximum value $\Omega_K=\sqrt{3.234 \,G \rho_0}$ is attained
for a model close to the maximum-mass model, with
$M=2.86 M_\odot (\rho_0/4.2\times10^{14}\,{\rm g\,cm^{-3}})^{-1/2}$.
In contrast to static
models, ISCO frequencies below 1.1 kHz are allowed---in fact, at the
canonical value $\rho_0 = 4.2\times 10^{14}\,{\rm g\,cm^{-3}}$, the
ISCO frequencies of rapidly rotating strange stars can be as low as
0.9 kHz for a $1.3 M_\odot$ strange star.  Hence, the presence of strange
stars in low-mass X-ray binaries cannot be excluded on the basis of
the currently observed frequencies of kHz QPOs, 
such as the cut-off frequency of 1066 Hz in 4U 1820-30.

\end{abstract}


\section{Introduction}

The discovery of millisecond variability in the flux of low-mass X-ray binaries
(LMXBs) has raised the prospect of constraining the properties of matter
at supranuclear densities, which is thought to make up the compact stellar
remnant in these sources. Conventionally, the compact objects is taken
to be a neutron star, and it has been shown
\cite{1997ApJ...480L..27K,1998ApJ...509L..37K} how the assumption that
the highest observed frequency in the X-ray flux
is the orbital frequency in the
innermost (marginally) stable orbit about the star
\cite{1985ApJ...297..548K,1986SvAL...12..117S,1990ApJ...358..538K}
 leads to significant constraints
on the equation of state of matter at such densities, 
as very few models of ultra-dense matter admit
neutron stars of mass high enough to allow maximum orbital frequencies
as low as the observed values in the quasi-periodic oscillations (QPOs)---for
instance,  the QPO frequency in 4U 1820-30 saturates at 1.07 kHz
\cite{1998ApJ...500L.171Z}.

Still, the evolutionary status of LMXBs and the nature of the
accreting compact object are not clear. It is known that the X-ray
bursters cannot be black holes, because their photospheric radius and
the temperature during the burst both tend to a definite value, thus showing
the presence of a stellar surface, which is also required to explain
the (type~I) X-ray bursts as thermonuclear explosions of accreted
material. The inferred radii (and masses) are consistent with models
of neutron stars, but it is possible that the compact object is a
``strange", i.e. quark, star \cite{1996ApJ...468..819C}. If it were,
at least in the sources 4U 1820-30 and 4U 1636-53, then the energy density
of self-bound quark matter at zero pressure would have
to have the unusually low value $\rho_0 <4.2\times 10^{14}\,{\rm
g\,cm^{-3}}$, if the maximum observed frequencies of the kHz QPOs
were the orbital frequencies in the ISCO about {\sl slowly}
rotating strange stars; and the observed
QPO frequencies could not be the orbital frequencies at the
{\sl surface} of such stars, for any value of $\rho_0$
 \cite{1999A&A...344L..71B}.

However, it seems likely that in these very old LMXBs, the compact star
has been spun up through accretion to very
 high\footnote{The frequency of $\sim 300\,$Hz
observed during some X-ray bursts is unlikely to be
the rotational frequency of the star,
as the long duration of the oscillation ($\sim 10\,$s), its harmonic content, 
and its variation during the burst, seem incompatible with the flux
modulation being caused by a ``hot spot'' on the stellar surface.}
 frequencies---a neutron
star would have to accrete only $\sim0.2M_\odot$ to attain angular
momentum $J=0.6GM^2/c$ \cite{1985ApJ...297..548K}. A strong magnetic
field could prevent rapid rotation, but strange stars may not
support such a field (Horvath 1999). 
We show that if LMXBs harbour rapidly rotating strange stars, the
constraints from kHz QPOs on the stellar mass and on $\rho_0$ are relaxed to
a surprising degree. We also discuss the rotational frequency of ``Keplerian"
models of strange stars, i.e. ones in which the rotation rate at the stellar
equator is equal to the orbital frequency for the star, at the same
equatorial radius.

\begin{figure}
\includegraphics[width=0.95\columnwidth]{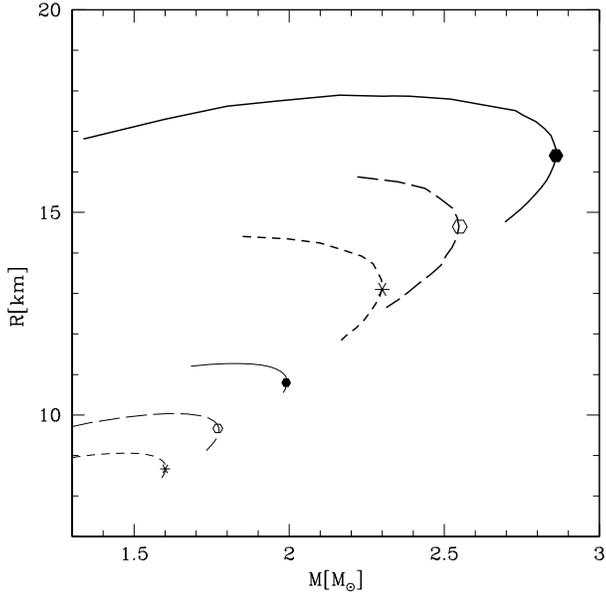}
\caption{ Radius vs. mass for strange stars (near their maximum mass).
Models of both static (thin lines) and maximally rotating (thick lines)
stars are shown for three values of the energy density of quark matter
at zero pressure, $\rho_0$:
$4.2\times 10^{14}\,{\rm g\, cm}^{-3}$ (continuous line)
$5.3\times 10^{14}\,{\rm g\, cm}^{-3}$ (long-dashed line),
$6.5\times 10^{14}\,{\rm g\, cm}^{-3}$ (dashed line).
The maximum-mass models for these values of $\rho_0$
are indicated by a filled circle, an open circle and
a star, respectively.
Note that the radius and mass scale as $\rho_0^{-1/2}$.
A large increase of the radius and maximum mass is evident as
the stellar rotation rate increases from zero to the equatorial
mass-shedding limit.     }
\end{figure}

\begin{figure}
\includegraphics[width=0.95\columnwidth]{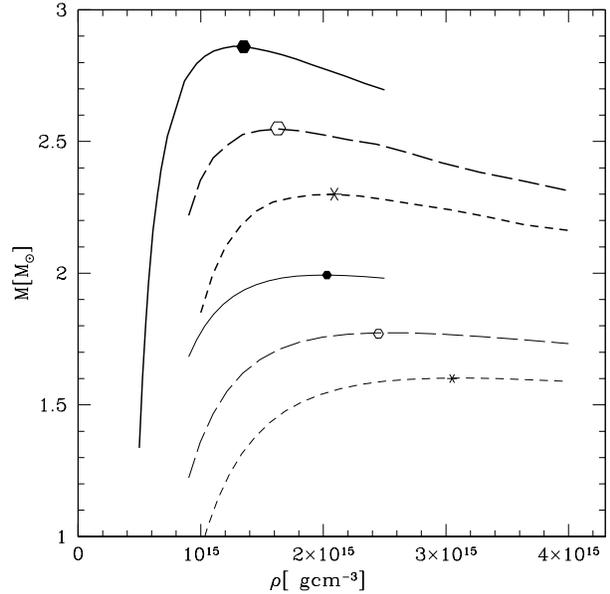}
\caption{Mass vs. central energy density for the same models as in Fig. 1.
For each sequence, a star is stable only at values of $\rho$
smaller than, approximately,  the one corresponding to the maximum mass.}
\end{figure}

\begin{table*}
\caption{Keplerian models of strange stars,
 for $\rho_0=4.2\times10^{14}{\rm g\,cm^{-3}}$.}

\begin{tabular}{cccccccccccccc}
$\rho_c$ & $M$ & $R$ & $\Omega$ & $\Omega_+$ & $h_+$ & $h_-$ & $J
$ &
$I$ & $r_p/r_e$ & $T/W$ & $z_p$ & $z_b$ & $-z_f$ \\
\hspace{-2mm}$10^{14}$g\,cm$^{-3}$ \hspace{-5mm}& \hspace{-3mm}$M_\odot$ &
 km & $10^3$s$^{-1}$ & $10^3$s$^{-1}$ & km & km 
& $GM_\odot^2/c$ & \hspace{-2mm}$10^{45}$g\,cm$^{2}$\hspace{-5mm}  
 &  &  &  &  \\
\hline\hline
$5.000 $
  & 
$1.336 $
  & 
$ 16.81 $
  & 
$7.049 $
  & 
$5.664 $
  & 
$1.86 $
  & 
$9.32 $
  & 
$2.349 $
  & 
$2.929 $
  & 
$ .3192 $ 
  & 
$ .2785 $ 
  & 
$ .2731 $ 
  & 
$ .8931 $ 
  & 
$ .2821 $ 
 \\
$5.250 $
  & 
$1.598 $
  & 
$ 17.29 $
  & 
$7.257 $
  & 
$5.877 $
  & 
$1.90 $
  & 
$ 11.2 $
  & 
$3.123 $
  & 
$3.782 $
  & 
$ .3365 $ 
  & 
$ .2709 $ 
  & 
$ .3279 $ 
  & 
$1.045 $
  & 
$ .2942 $ 
 \\
$5.500 $
  & 
$1.800 $
  & 
$ 17.62 $
  & 
$7.432 $
  & 
$6.051 $
  & 
$1.87 $
  & 
$ 12.7 $
  & 
$3.760 $
  & 
$4.447 $
  & 
$ .3500 $ 
  & 
$ .2645 $ 
  & 
$ .3739 $ 
  & 
$1.176 $
  & 
$ .3048 $ 
 \\
$5.750 $
  & 
$1.963 $
  & 
$ 17.75 $
  & 
$7.591 $
  & 
$6.217 $
  & 
$1.86 $
  & 
$ 13.9 $
  & 
$4.299 $
  & 
$4.978 $
  & 
$ .3630 $ 
  & 
$ .2590 $ 
  & 
$ .4145 $ 
  & 
$1.294 $
  & 
$ .3113 $ 
 \\
$6.105 $
  & 
$2.161 $
  & 
$ 17.89 $
  & 
$7.786 $
  & 
$6.389 $
  & 
$1.88 $
  & 
$ 15.6 $
  & 
$5.003 $
  & 
$5.647 $
  & 
$ .3750 $ 
  & 
$ .2534 $ 
  & 
$ .4676 $ 
  & 
$1.453 $
  & 
$ .3199 $ 
 \\
$6.500 $
  & 
$2.304 $
  & 
$ 17.87 $
  & 
$7.979 $
  & 
$6.582 $
  & 
$1.84 $
  & 
$ 16.7 $
  & 
$5.474 $
  & 
$6.030 $
  & 
$ .3906 $ 
  & 
$ .2465 $ 
  & 
$ .5121 $ 
  & 
$1.589 $
  & 
$ .3260 $ 
 \\
$6.600 $
  & 
$2.336 $
  & 
$ 17.87 $
  & 
$8.012 $
  & 
$6.615 $
  & 
$1.84 $
  & 
$ 16.9 $
  & 
$5.590 $
  & 
$6.131 $
  & 
$ .3930 $ 
  & 
$ .2457 $ 
  & 
$ .5221 $ 
  & 
$1.619 $
  & 
$ .3270 $ 
 \\
$6.750 $
  & 
$2.388 $
  & 
$ 17.86 $
  & 
$8.080 $
  & 
$6.678 $
  & 
$1.84 $
  & 
$ 17.4 $
  & 
$5.788 $
  & 
$6.296 $
  & 
$ .3969 $ 
  & 
$ .2441 $ 
  & 
$ .5393 $ 
  & 
$1.674 $
  & 
$ .3291 $ 
 \\
$7.275 $
  & 
$2.520 $
  & 
$ 17.79 $
  & 
$8.280 $
  & 
$6.864 $
  & 
$1.82 $
  & 
$ 18.6 $
  & 
$6.258 $
  & 
$6.643 $
  & 
$ .4094 $ 
  & 
$ .2387 $ 
  & 
$ .5875 $ 
  & 
$1.829 $
  & 
$ .3348 $ 
 \\
$8.668 $
  & 
$2.730 $
  & 
$ 17.51 $
  & 
$8.708 $
  & 
$7.229 $
  & 
$1.79 $
  & 
$ 20.5 $
  & 
$7.007 $
  & 
$7.072 $
  & 
$ .4312 $ 
  & 
$ .2285 $ 
  & 
$ .6835 $ 
  & 
$2.153 $
  & 
$ .3458 $ 
 \\
$9.023 $
  & 
$2.753 $
  & 
$ 17.40 $
  & 
$8.803 $
  & 
$7.327 $
  & 
$1.76 $
  & 
$ 20.7 $
  & 
$7.055 $
  & 
$7.043 $
  & 
$ .4367 $ 
  & 
$ .2257 $ 
  & 
$ .6999 $ 
  & 
$2.207 $
  & 
$ .3474 $ 
 \\
$9.658 $
  & 
$2.796 $
  & 
$ 17.23 $
  & 
$8.958 $
  & 
$7.459 $
  & 
$1.75 $
  & 
$ 21.1 $
  & 
$7.180 $
  & 
$7.045 $
  & 
$ .4437 $ 
  & 
$ .2221 $ 
  & 
$ .7290 $ 
  & 
$2.309 $
  & 
$ .3501 $ 
 \\
$ 10.33 $
  & 
$2.823 $
  & 
$ 17.06 $
  & 
$9.103 $
  & 
$7.582 $
  & 
$1.74 $
  & 
$ 21.3 $
  & 
$7.233 $
  & 
$6.983 $
  & 
$ .4504 $ 
  & 
$ .2186 $ 
  & 
$ .7535 $ 
  & 
$2.394 $
  & 
$ .3523 $ 
 \\
$ 11.06 $
  & 
$2.844 $
  & 
$ 16.89 $
  & 
$9.245 $
  & 
$7.699 $
  & 
$1.73 $
  & 
$ 21.5 $
  & 
$7.259 $
  & 
$6.901 $
  & 
$ .4562 $ 
  & 
$ .2154 $ 
  & 
$ .7760 $ 
  & 
$2.474 $
  & 
$ .3544 $ 
 \\
$ 11.84 $
  & 
$2.854 $
  & 
$ 16.69 $
  & 
$9.385 $
  & 
$7.823 $
  & 
$1.72 $
  & 
$ 21.6 $
  & 
$7.232 $
  & 
$6.772 $
  & 
$ .4625 $ 
  & 
$ .2119 $ 
  & 
$ .7953 $ 
  & 
$2.541 $
  & 
$ .3555 $ 
 \\
$ 12.67 $
  & 
$2.861 $
  & 
$ 16.52 $
  & 
$9.522 $
  & 
$7.933 $
  & 
$1.71 $
  & 
$ 21.7 $
  & 
$7.211 $
  & 
$6.656 $
  & 
$ .4672 $ 
  & 
$ .2090 $ 
  & 
$ .8140 $ 
  & 
$2.608 $
  & 
$ .3573 $ 
 \\
$ 13.56 $
  & 
$2.860 $
  & 
$ 16.33 $
  & 
$9.648 $
  & 
$8.047 $
  & 
$1.69 $
  & 
$ 21.7 $
  & 
$7.124 $
  & 
$6.489 $
  & 
$ .4727 $ 
  & 
$ .2056 $ 
  & 
$ .8271 $ 
  & 
$2.652 $
  & 
$ .3579 $ 
 \\
$ 14.51 $
  & 
$2.852 $
  & 
$ 16.14 $
  & 
$9.782 $
  & 
$8.153 $
  & 
$1.69 $
  & 
$ 21.7 $
  & 
$7.038 $
  & 
$6.323 $
  & 
$ .4773 $ 
  & 
$ .2026 $ 
  & 
$ .8415 $ 
  & 
$2.702 $
  & 
$ .3589 $ 
 \\
$ 15.53 $
  & 
$2.842 $
  & 
$ 15.97 $
  & 
$9.909 $
  & 
$8.256 $
  & 
$1.68 $
  & 
$ 21.6 $
  & 
$6.937 $
  & 
$6.153 $
  & 
$ .4812 $ 
  & 
$ .1998 $ 
  & 
$ .8532 $ 
  & 
$2.741 $
  & 
$ .3600 $ 
 \\
$ 16.63 $
  & 
$2.829 $
  & 
$ 15.78 $
  & 
$ 10.03 $
  & 
$8.363 $
  & 
$1.67 $
  & 
$ 21.5 $
  & 
$6.829 $
  & 
$5.980 $
  & 
$ .4852 $ 
  & 
$ .1970 $ 
  & 
$ .8638 $ 
  & 
$2.777 $
  & 
$ .3606 $ 
 \\

\end{tabular}
\end{table*}

\section{Strange stars}

Quark stars are likely to exist if the ground state of matter at large 
atomic number is in the form of a quark fluid, which would then
necessarily be composed of about equal numbers of up, down, and strange
quarks \cite{1971Bodmer}. Today, such matter is called strange matter. Its
thermodynamic properties have been discussed in detail by Farhi
and Jaffe (1984) in the context of quantum chromodynamics.
The first relativistic model of stars composed of quark matter
was computed by \cite{1976Natur.259..377B}. The cosmological
consequences of the presumed existence of strange matter were 
first discussed in detail by \cite{1984Witten}, who also showed that
the maximum mass 
of a static strange star scales as $\rho_0^{-1/2}$ and is $2M_\odot$
for $\rho_0=4\times 10^{14}\,{\rm g\,cm^{-3}}$. Detailed models
of strange stars have been constructed by
\cite{1986ApJ...310..261A}
and \cite{1986A&A...160..121H}.

Astrophysical implications of these ideas are not yet clear.
It has been pointed out that young, glitching radio pulsars
are probably neutron stars \cite{1987S&T....74..580A}, and that strange stars
are unlikely to be present in Hulse-Taylor type binaries, as their
coalescence may lead to dispersal of nuggets of quark matter which
would have precluded the formation of young neutron stars in the
Galaxy
(Madsen 1988, Caldwell and Friedman 1991). However, there seems
to be no objection to millisecond pulsars or the compact objects in
LMXBs being strange stars
\cite{1994A&A...286L..17K,1996ApJ...468..819C},
and it has even been suggested (Madsen 1999)
that millisecond pulsars can be formed
directly in supernovae, if they are strange stars [unlike neutron stars,
whose rotation rate would be quickly damped by the r-mode instability
\cite{1999ApJ...510..846A}, \cite{LOM98}, \cite{1999A&A...341..110K}].

Our work is informed by the question whether the presence of strange
stars in LMXBs can be excluded on the basis of the observed timing
properties of these sources.\footnote{In the conversion of accreted matter
to quark matter, substantial energy will be released at the base of the crust.
However, since the mass accretion rate in LMXBs
is only inferred from the X-ray luminosity, there is no direct way of
differentiating this from the gravitational
binding energy release,
if the conversion of matter is proceeding at a quasi-steady rate.}
 Specifically, it has been asked whether
the  observed frequency of the kHz QPOs in sources such as 4U 1820
-30  (1.07 kHz) is not too low to be compatible with the maximum
mass of strange stars \cite{1999A&A...344L..71B}. For this reason, in
constructing our models of strange stars, we have focussed on an equation
of state which yields the largest masses of static strange star models:
$$ P = (\rho - \rho_0)/3. \eqno (1)$$
 We have
also investigated the more general case, where the factor of 1/3 in
eq. (1) is replaced by a different positive constant, $a \le 1$.

\section{Keplerian models of strange stars}

We have computed exact numerical models of strange stars in general
relativity using the \cite{SF95} code (see Stergioulas 1998
for a description).
In this code, the equilibrium models are obtained following the KEH
(Komatsu et al. 1989) method, in which the field equations are converted
to integral equations using appropriate Green's functions.

\begin{figure}
\includegraphics[width=0.95\columnwidth]{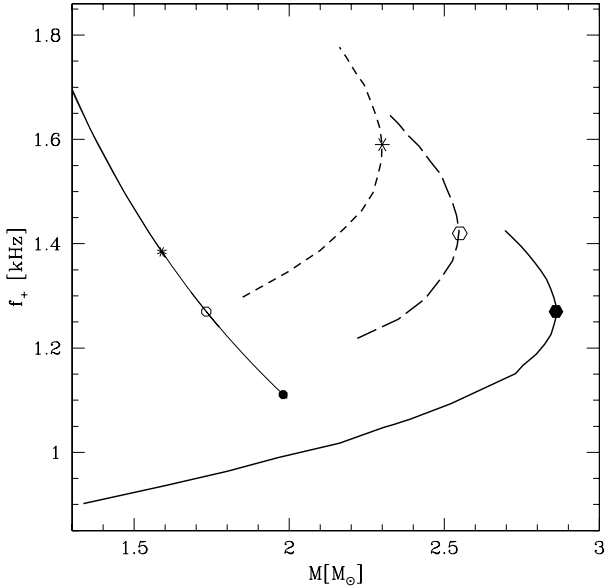}
\caption{The frequency of the co-rotating innermost stable circular orbit
as a function of mass for static models (thin, continuous line)
and for strange stars rotating at the equatorial mass-shedding
limit (thick lines, in the style of Fig. 1). For the static models,
$f_+ = 2198\, {\rm Hz}(M_\odot/M)$, and the minimum ISCO frequency
corresponds to the maximum mass, denoted by a filled circle, an empty circle,
and a star, respectively for
$\rho_0/(10^{14}g\,cm^{-3})=$ 4.2, 5.3, and 6.5.
Note that the ISCO frequencies for rapidly rotating strange stars
can have much lower values, and $f_+<1\,$kHz can be achieved
for strange stars of fairly modest mass, e.g. $1.4M_\odot$,
if the star rotates close to the equatorial mass-shedding limit.   
}
\end{figure}

\begin{figure}
\includegraphics[width=0.95\columnwidth]{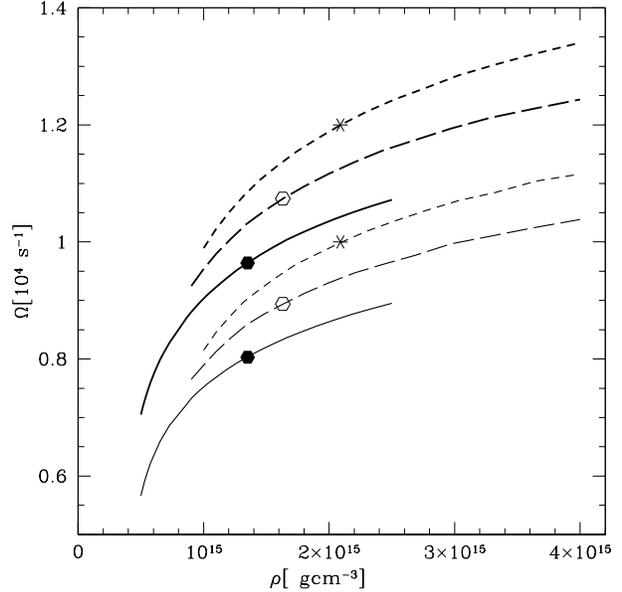}
\caption{ The angular frequency of strange stars (thick lines) rotating at
the equatorial mass-shedding limit, and of their co-rotating ISCOs
(thin lines), as a function of central energy density.
The style of lines corresponds to the same values of $\rho_0$
as in Fig. 1.}
\end{figure}

The detailed expected properties of strange stars depend on the
adopted theory of interactions. All models presented here were
constructed using eq. (1) for the equation of state. For our models
of rotating stars, we find that mass and radial quantities (e.g.,
the stellar radius and the height
of the ISCO above it) scale as $\rho_0^{-1/2}$, just as for the static stars,
while the frequencies scale as $\rho_0^{1/2}$.
For the more general e.o.s.
$P=a (\rho -\rho_0)$, we confirm the approximate scalings of
\cite{1990ApJ...355..241L}
to within 9 \%, that is, we find that for our Keplerian models,
the maximum stellar mass scales as $a^{1/2}$,
the stelar radius scales as $a^{1/4}$,
and the rotation rate of the star scales as $a^{-1/8}$. 
Thus, between the scalings with $\rho_0$ and with $a$, the
numerical results presented here
can at once be extended to the general e.o.s. of strange matter.

In Figs. 1 through 3, we present the mass, radius and the ISCO
frequencies in our Keplerian models, for three values of $\rho_0$,
and compare them with the static models. In Fig. 4 we present the
ISCO angular frequencies as a function of the central energy density
of the strange star, and also exhibit the (larger) angular frequency
of the star itself. Note, that because of the scalings with energy
density, the maximal rotation rate of a strange star (which is very
close to the rotation rate of the maximum-mass model in Table 1) is
described by the simple formula $\Omega_K= \sqrt{3.234 \,G\rho_0}$,
where $G$ is the gravitational constant.

Table 1 presents in detail, the various stellar parameters obtained
in our calculation for Keplerian strange stars, modeled with eq. (1)
for the value $\rho_0=4.2 \times 10^{14} \,{\rm g\,cm^{-3}}$.
Note that in all cases, the co-rotating ISCO is above the stellar
surface and, as exhibited in Fig. 3, the ISCO frequencies are much
lower for the Keplerian models than for static models (at fixed stellar mass)
and differ considerably from their lowest-order slow-rotation approximation.
The significant departure from the slow-rotation result is explained by the
unusually large oblateness of rapidly rotating strange stars, and the
fact that the ISCO frequency and height depend not only on the angular 
momentum, but also on the stationary quadrupole moment, in rapidly 
rotating stars \cite{SS98}.

\section{Conclusions}

We have calculated exact models of rotating strange stars---these
computationss are in excellent agreement (Table 2)
with the very recent results
obtained by a highly accurate code based on spectral methods
\cite{Gurgulon}.
 We found the scalings of $M$, $R$, $\Omega$ with the parameters
$a$, and $\rho_0$ in the equation of state of self-bound quark matter
$P=a(\rho -\rho_0)$. In addition, we calculate the
innermost stable orbits and find that,
unlike for static models, for strange stars rotating
at the equatorial mass-shedding limit the orbital
frequencies can extend to values below $1.1\,$kHz.
For the same models the radius of the
ISCO is about 11\% larger than the circumferential stellar radius,
independently of the central density, or the value of $\rho_0$.

Our results show that the highest observed QPO frequencies in low-mass
X-ray binaries (such as 1.07 kHz in 4U 1820-30) could be the orbital
frequencies in the innermost stable circular orbit about strange stars,
if only the stars are rotating sufficiently rapidly (as is expected
in these old accreting sources).
Thus, the compact objects in LMXBs could, in principle,
 be rapidly rotating strange stars. Further conclusions about the nature
of LMXBs and of the kHz QPOs would be possible, if either the mass
or the rotational period of the accreting stellar remnant were known.

\newcommand{\titen}[1]{\times 10^{#1}}
\begin{table}
\caption{Comparison of two codes.}
\begin{tabular}{lllll}
       &  &This work             & Gourgoulhon & diff. \\
       &  &                      &~~~~et~al.  &   [\%]       \\
	 \hline\hline
$\rho_c$& $10^{15}$g\,cm$^{-3}$  &  $1.261$ &~~~~$1.261$  &  \\
$M$     & $M_\odot$ &  $2.83392$           &~~~~$2.831 $         & $0.1$ \\
$R$     & km & $16.4252$           &   $~~~16.54$           & $0.7$ \\
$\Omega$&$10^3$s$^{-1}$& $9.56261$           &~~~~$9.547$         & $0.2$ \\
$\Omega_K$&$10^3$s$^{-1}$& $9.57637$           &~~~~$9.547$      & $0.3$ \\
$T/W$    & & $0.210091$          &~~~~$0.210$            & $0.04$\\
$cJ/GM_\odot^2$&&
            $7.09331$           &~~~~$7.084$            & $0.1$ \\
$I$   & $10^{45}$g\,cm$^2$&
            $6.51906$           &~~~~$6.534$            & $0.2$ \\
$z_p$   &&  $0.807995$          &~~~~$0.8070$           & $0.1$ \\
$z_b$   &&  $2.58820$           &~~~~$2.584$            & $0.2$ \\
$-z_f$  &&  $0.356272$         &~~~~$0.3608$          & $1.2$ \\
$r_p/r_e$ &&  $0.466000$          &~~~~$0.4618$           & $0.9$
\end{tabular}
\end{table}

\bigskip\bigskip

This research was supported in part by the NSF grant PHY94-07194.

\newcommand{\apjl}{ApJ Let.}
\newcommand{\apj}{ApJ}
\newcommand{\mnras}{MNRAS}
\newcommand{\nat}{Nature}
\newcommand{\apjs}{ApJ Supp.}
\newcommand{\aap}{A\&A}

\bibliography{aamnem99,bkhz}
\bibliographystyle{aabib99}

\end{document}